\documentclass[11pt]{article}
\usepackage{fullpage}

\usepackage[utf8]{inputenc}
\usepackage{multirow}

\usepackage[shortlabels]{enumitem}
\usepackage{amsmath}
\usepackage{amssymb}
\usepackage{amsthm}
\usepackage[dvipsnames]{xcolor}
\usepackage{mathtools}
\usepackage[utf8]{inputenc}
\usepackage{amsfonts}
\usepackage{graphicx}
\usepackage{mathrsfs}
\usepackage{physics}
\usepackage{complexity}
\usepackage{soul}
\usepackage[ruled,linesnumbered]{algorithm2e}
\usepackage[T1]{fontenc}
\usepackage{hyperref}
\usepackage[capitalize]{cleveref}
\usepackage[hyperpageref]{backref}
\usepackage{bbm}

\usepackage{todonotes}
\usepackage[
	lambda,
	operators,
	advantage,
	sets,
	adversary,
	landau,
	probability,
	notions,	
	logic,
	ff,
	mm,
	primitives,
	events,
	complexity,
	asymptotics,
	keys]{cryptocode}

\usepackage{tikz}
\usetikzlibrary{patterns}

\newcommand{\calT}{\mathcal{T}}

\newcommand{\calW}{\mathcal{W}}
\newcommand{\calO}{\mathcal{O}}
\newcommand{\calR}{\mathcal{R}}

\newcommand{\from}{\leftarrow}
\renewcommand{\E}{\mathop{{}\mathbb{E}}}
\newcommand{\Model}{\mathsf{Model}}
\newcommand{\RModel}{\overline{\mathsf{Model}}}

\newcommand{\Detect}{\mathsf{Detect}}
\newcommand{\prompt}{\textsc{prompt}}
\newcommand{\Wat}{\mathsf{Wat}}
\newcommand{\Setup}{\mathsf{Setup}}
\newcommand{\done}{\texttt{done}}
\newcommand{\empH}{H_e}

\newtheorem{theorem}{Theorem}

\newtheorem{definition}{Definition}
\newtheorem{lemma}{Lemma}

\title{Undetectable Watermarks for Language Models}
\author{Miranda Christ\thanks{Equal contribution.}\\ \small Columbia University 
        \and 
        Sam Gunn$^*$\thanks{Supported by a Google PhD Fellowship.}\\ \small UC Berkeley
        \and 
        Or Zamir$^*$\\ \small Princeton University}
\begin{document}

\maketitle

\begin{abstract}
Recent advances in the capabilities of large language models such as GPT-4 have spurred increasing concern about our ability to detect AI-generated text. 
Prior works have suggested methods of embedding watermarks in model outputs, by \emph{noticeably} altering the output distribution. We ask: Is it possible to introduce a watermark without incurring \emph{any detectable} change to the output distribution?

To this end we introduce a cryptographically-inspired notion of undetectable watermarks for language models. 
That is, watermarks can be detected only with the knowledge of a secret key; without the secret key, it is computationally intractable to distinguish watermarked outputs from those of the original model. In particular, it is impossible for a user to observe any degradation in the quality of the text. Crucially, watermarks should remain undetectable even when the user is allowed to adaptively query the model with arbitrarily chosen prompts. We construct undetectable watermarks based on the existence of one-way functions, a standard assumption in cryptography.
\end{abstract}

\thispagestyle{empty}
\newpage
{\small\tableofcontents}
\newpage

\section{Introduction}

With the rise in the use of artificial models that churn out human-like text, there's also an increase in the potential for misuse. Imagine a student employing a language model to effortlessly write her ``Machine Learning 101'' homework or conjuring up tear-jerking emails to beg professors for easier exams. That's when the need arises to distinguish between texts penned by a language model and those crafted by human hands. 
The go-to method of employing a heuristic test to determine if a text was AI-generated, however, grows increasingly fragile as large language models (LLMs) advance. Even the cutting-edge detectors, like GPTZero~\cite{gptzero}, can be outsmarted with cleverly crafted prompts.

Ultimately, as LLM outputs move closer to becoming identical to human-generated text, this approach becomes hopeless. 
It is already very hard to tell, for instance, that the previous paragraph was written by such a model.
To overcome this problem, it is reasonable to consider intentionally modifying the model to embed \emph{watermarks} into the text.
Recent work of~\cite{kgw} introduced such watermarks in the context of LLMs.
However, existing watermarking schemes come with a cost: To plant a useful watermark, the distribution of texts the model generates has to be noticeably changed. In fact, for existing schemes it is possible for the \emph{user} to distinguish between outputs of the original model and of the watermarked one, and it is hence possible that the quality of text degrades.

We show how to plant watermarks with the following properties, stated informally, in any LLM.
\begin{enumerate}
    \item\label{property:undetect} (Undetectability) It is computationally infeasible to distinguish between the original and the watermarked models, even when the user is allowed to make many adaptive queries. In particular, the quality of generated text remains identical.
    \item\label{property:complete} (Completeness) There is a secret key that enables efficient detection of responses from the watermarked model, as long as ``enough randmoness'' was used to generate the response. The detection works even when presented with only a contiguous sub-string from the response, and it doesn't require any other information.
    \item\label{property:sound} (Soundness) Any text generated independently from the secret key has a negligible chance of being detected as watermarked.
\end{enumerate}
We note that the existence of a secret key is necessary, as otherwise Properties~(\ref{property:undetect}) and~(\ref{property:complete}) would directly contradict each other. However, in practice the secret key can be published if one wishes; Property~(\ref{property:undetect}) still ensures that the quality of the text is imperceptibly changed for all uses not involving the secret key.


An important aspect of our construction is that Properties~(\ref{property:undetect}) and~(\ref{property:sound}) will \emph{always} hold, for any LLM with any choice of parameters, and without making any assumptions on the text.
Our scheme is the first to have these properties, and we argue that they are completely crucial. First, the creator of a state-of-the-art LLM is unlikely to intentionally degrade the quality of their model, making Property~(\ref{property:undetect}) necessary for any practical watermarking scheme. As the quality and versatility of LLMs has reached such high levels, any noticeable change due to the watermark is liable to have adverse side-effects in some situations. Second, falsely accusing humans of using LLMs to generate their texts should be completely unacceptable. When heuristics are used for detection, this will always be a possibility --- indeed, instances of such false accusations against students have already made news headlines \cite{wapo, usatoday}, and concerningly, false accusations are more common for non-native English writers \cite{liang2023gpt}. Property~(\ref{property:sound}) in our construction rigorously guarantees that natural text will not be detected as watermarked.

Of course, a watermark is only useful if it can be detected with the secret key. If the model has a deterministic response to some prompt, then we should not be able to embed the watermark in that response (as any change to the output would necessarily be detectable). For Property~(\ref{property:complete}), we therefore need to assume that enough ``randomness'' was used in the generation of the specific text we are considering.
We introduce a formal notion that we call \emph{empirical entropy}, and show that this condition is necessary.
Our detection algorithm works when it is given text containing any consecutive sub-string with enough empirical entropy from an output of the model.

Primary contributions of this work include the formal definition and construction of \emph{undetectable watermarks}, and the notion of \emph{empirical entropy} that quantifies the randomness used in the generation of a specific output. These definitions and our results appear in \Cref{sec:model}.

\subsection{Related Work}\label{subsec:recent}
Approaches for detecting AI-generated text largely fall into two categories. \emph{Watermarking schemes} alter the output of a language model in a way that a corresponding detection algorithm can identify. \emph{Post-hoc detectors} leave the output of the model unchanged and instead identify AI-generated text using existing differences between natural language and the model's output.

\paragraph{Post-hoc detectors.}
The simplest post-hoc detectors use natural heuristics to distinguish between human- and AI-generated text.
These heuristics include relative entropy scoring \cite{lavergne2008detecting}, perplexity \cite{beresneva2016computer}, and other statistical methods \cite{gehrmann2019gltr}; see \cite{beresneva2016computer} for a survey of such methods.
Other post-hoc detectors (e.g., \cite{zellers2019defending, detectgpt, gptzero, openaidetector}) are themselves models, specifically trained for this binary classification task.
Unfortunately, these heuristic and model-based methods lack formal guarantees, and it's possible to train a model to transform AI-generated text in a way that evades them; see, e.g., \cite{krishna2023paraphrasing, DBLP:journals/corr/abs-2303-11156}.
For example, \cite{krishna2023paraphrasing} trains a model to paraphrase text output by language models, fooling common post-hoc detectors such as GPTZero \cite{gptzero}, DetectGPT \cite{detectgpt}, and the detector developed by OpenAI \cite{openaidetector}. 
Furthermore, simple tricks such as instructing the model in the prompt to write a response that evades a detector, or varying the model's parameters (e.g., increasing the temperature and frequency/presence penalties for GPT-4), fool \cite{gptzero}.
\cite{chakraborty2023possibilities} prove that as AI-generated text more closely resembles natural text, post-hoc detectors will need longer text samples. 

See \cite{jawahar2020automatic} for more comprehensive background on post-hoc detection of AI-generated text and attacks.
    



\paragraph{Language watermarking schemes.}
Several schemes (e.g., \cite{adversarial, paraphraser, yoo2023robust, deeptextmark}) involve using an ML model in the watermarking algorithm itself. 
\cite{adversarial, deeptextmark} work by taking a passage of text and using a model to produce a semantically similar altered passage.
By nature of using machine learning, these constructions have no formal guarantees and rely on heuristic arguments for undetectability, soundness, and correctness.

In a recent work, \cite{kgw} presented the first watermarking scheme for LLMs with any formal guarantees. They showed that a watermark can be planted in outputs with large enough entropy (with a definition different than ours, yet morally similar). However, their watermarking scheme crucially \emph{changes} the distribution of generated texts and uses this change to detect the watermark. They bound the difference between the original distribution and the distribution of their watermarked model, using a quantity called \emph{perplexity}.
In contrast, in our work the original and watermarked output distributions are completely indistinguishable. 

The authors of this paper are also aware of an ongoing watermarking project of~\cite{scottblog}, which he mentions in his blog.
It appears that in this project, the guarantee is that the two distributions will be indistinguishable, but only as long as no two output texts are seen that share a common substring of a certain length. 
In contrast, our construction guarantees undetectability without any assumption on the texts or the model.
In particular we allow the distinguisher to make adaptive queries with arbitrary prompts, so it may force the model to return outputs that share long parts with each other. 

Our scheme, as well as those of \cite{kgw} and \cite{scottblog} are vulnerable to simple attacks such as the ``emoji attack''\footnote{\url{https://twitter.com/goodside/status/1610682909647671306}} discussed in \cite{kgw}. We discuss attacks on watermarking schemes further in \Cref{subsec:attacks}.

\paragraph{Steganography.}
Steganography is the study of encoding a hidden message into a given channel (e.g., natural language or an image) such that a recipient possessing a key can read the message but an eavesdropping adversary cannot determine whether a message is present.
\cite{steg} defines security of a steganography scheme against a chosen hiddentext attack (CHA) as the requirement that an adversary cannot determine whether a given oracle is for the original distribution or the distribution embedded with a hiddentext.
Porting this definition over to the watermark setting, one would obtain undetectability. However, the corresponding steganography schemes we are aware of would not obtain undetectability \emph{without making assumptions about the entropy of the channel}. Crucially, in our setting where prompts for the language model are adversarially chosen, we want to retain undetectability even if the adversary submits a prompt with a deterministic response.

Part of the reason for this difference stems from the access that the watermark and encoding algorithms have to the channel (or the distribution of the language model's output).
In steganography, the encoding algorithm has only oracle access to the channel.
In our language model setting, the watermark algorithm receives from the model a full description of the probability distribution $p_i$ for each token. 
Because of this limited access, steganography schemes largely either rely on assumptions about the entropy of the channel (e.g., \cite{steg}) or lose security when the channel has low entropy (e.g., \cite{dedic2009upper}).
Our watermark is undetectable \emph{regardless} of the entropy of the text.
We achieve this guarantee exactly by using the watermark algorithm's access to the distributions $p_i$. 
Using this knowledge, the algorithm is able to compute the \emph{empirical entropy} of its output thus far. Once the empirical entropy is sufficiently high, it uses the output as a random seed for a PRF used to embed the watermark in subsequent tokens.
Importantly, our algorithm alters the output distribution only once it has collected enough entropy; in the steganography schemes, the encoding algorithm with only oracle access does not know when this has happened.

This complete knowledge of each token distribution $p_i$ separates the problem of watermarking language models from watermarking more generally. 
One prior steganography scheme for language models, \cite{meteor}, operates in our regime where the encoding algorithm has access to each $p_i$.
However, it assumes that the decoding algorithm has access to each $p_i$ as well. 
This is unrealistic for watermarking, as the probability distributions $p_i$ for the output of the model on some prompt depend on that prompt.
A watermark detection algorithm, which receives only the output and not the prompt, does not know the distributions $p_i$. 
In practice, it would be unrealistic to assume that a detector trying to determine whether an essay was generated by a language model would also be given the prompt used to generate it.

\paragraph{Watermarking.}
The field of digital watermarking focuses on the problem of covertly planting a signal in a medium (e.g., an image or text) such that it can be detected by an algorithm. 
\cite{atallah2001natural, atallah2003natural} present some of the first watermarking schemes for NLP-generated text, though their schemes rely on a syntactic tree structure that was present in NLP models at that time but no longer used today.
\cite{hopper2007weak} formally defines watermarking and desired properties, though these definitions are not tailored to language models. 

\paragraph{Other related work.}
Recently,~\cite{goldwasser2022planting} used a cryptographic construction to embed undetectable backdoors into neural networks. In both their work and ours, cryptographic notions of indistinguishability are used in the context of machine-learning, rather than empirical notions.


\subsection{Organization of the Paper}
In Section~\ref{sec:model} we formally define our model, introduce the notions of empirical entropy and undetectable watermarks, and outline our results.
In Section~\ref{sec:inefficient} we sketch a simple watermarking scheme that achieves undetectability, but falls short of our main scheme in other respects.
In Section~\ref{sec:wm} we construct our undetectable watermarks with strong completeness and soundness guarantees. We include an overview of these constructions in Section~\ref{subsec:overview}.
In Section~\ref{sec:nec_assump} we discuss the necessity of the assumptions we make.
In Section~\ref{sec:remove} we discuss possible methods of removing watermarks from texts. In particular, we prove that it is impossible to create undetectable watermarks that are completely unremovable, under certain assumptions.
In Section~\ref{sec:open} we summarize and discuss open problems.

\section{Modeling the Problem}\label{sec:model}
\subsection{Preliminaries}\label{subsec:prelim}
\paragraph{Notation.}
Let $\secpar$ denote the security parameter. 
A function $f$ of $\secpar$ is \emph{negligible} if $f(\secpar) \in O(\frac{1}{\poly(\secpar)})$ for every polynomial $\poly(\cdot)$.
We denote $f(\secpar) \leq \negl$ to mean that $f$ is negligible.
For a vector or sequence of tokens $s = (s_1, \ldots, s_{\abs{s}})$ and positive integers $b \geq a$, let $s[a:b]$ denote $(s_a, \ldots, s_b)$.
We use $\log(x)$ to denote the logarithm base 2 of $x$, and $\ln(x)$ to denote the natural logarithm of $x$.
For integer $n > 0$, we define $[n] := \{1, \dots, n\}$. For integers $n \ge k > 0$, we define $[k,n] := \{k, \dots, n\}$.

\paragraph{Pseudorandom function (PRF).} Let $\mathcal{F} = \{F_{\sk} : \{0,1\}^{\ell_1(\secpar)} \to \{0, 1\}^{\ell_2(\secpar)} | \sk \in \{0,1\}^\secpar\}$ be a family of functions. 
$\mathcal{F}$ is a PRF if $F_\sk$ is efficiently computable and for all p.p.t. distinguishers $D$,
\[\left| \Pr_{\sk \gets \{0,1\}^\secpar}\left[D^{F_\sk(\cdot)}(1^\secpar) = 1\right] - \Pr_{f} \left[D^{f(\cdot)}(1^\secpar) = 1 \right]\right| \leq \negl\]
where $f$ denotes a random function from $\{0,1\}^{\ell_1(\secpar)}$ to $\{0,1\}^{\ell_2(\secpar)}$.
PRFs are a standard cryptographic primitive equivalent to one-way functions and can be constructed from standard assumptions \cite{goldreich1986construct, haastad1999pseudorandom}.

\subsection{Language Models}
We loosely follow \cite{kgw} in our definition of a \emph{language model}. We will often refer to language models simply as \emph{models}.
\begin{definition}\label{def:model}
A language model $\Model$ over token set~$\cal{T}$ is a deterministic algorithm that takes as input a prompt $\prompt$ and tokens previously output by the model $x=(x_1, \ldots, x_{i-1})$, and outputs a probability distribution $p_i = \Model(\prompt, x)$ over $\cal{T}$.
\end{definition}

A language model~$\Model$ is used to generate text as a response to a prompt by iteratively sampling from the returned distribution until a special terminating token~$\done \in \cal{T}$ is drawn.
\begin{definition}\label{def:modelresp}
    A language model's \emph{response} to a prompt~$\prompt$ is a random variable~$\RModel(\prompt)\in \mathcal{T}^\star$ that is defined algorithmically as follows.
    We begin with an empty list of tokens~$x=()$. As long as the last token in~$x$ is not~$\done$, we draw a token~$x_i$ from the distribution~$\Model(\prompt,x)$ and append it to~$x$. Finally, we set~$\RModel(\prompt) = x$.
\end{definition}
Throughout the text we will make use of a security parameter $\lambda$. We will assume that our model never outputs text of length super-polynomial in $\lambda$. (For OpenAI's language models, there is actually a fixed limit to the length of generated text.)

\subsection{Entropy and Empirical Entropy}
Let $\log(x)$ denote the logarithm base 2 of $x$.
For a probability distribution~$D$ over elements of a finite set~$X$, we define the Shannon \emph{entropy} of~$D$ as 
\[H(D) = \E_{x\sim D}[-\log D(x)],\]
where~$D(x)$ is the probability of~$x$ in the distribution~$D$.
The \emph{empirical entropy} of~$x$ in~$D$ is simply~$-\log D(x)$.
The expected empirical entropy of~$x\sim D$ is exactly~$H(D)$.
Intuitively, the empirical entropy of~$x$ (with respect to~$D$) is the number of random bits that were required to draw~$x$ out of the distribution~$D$.
The entropy~$H(D)$ is thus the expected number of random bits needed to draw an element out of the distribution~$D$.

We thus define the empirical entropy of a model's response as follows. 
\begin{definition}\label{def:empentropy}
    For a language model~$\Model$, a prompt~$\prompt$, and a possible response~$x\in\calT^\star$, we define the \emph{empirical entropy} of~$\RModel$ responding with~$x$ to~$\prompt$ as
    \[
        \empH(\Model,\prompt,x) := -\log \Pr\left(\RModel\left(\prompt\right)=x\right).
    \]
\end{definition}

We next generalize the definition of empirical entropy from whole outputs to \emph{substrings} out of a model's output. Intuitively, we want to measure how much entropy was involved in the generation of a particular contiguous substring of the output.

\begin{definition}\label{def:subseq-empentropy}
    For a language model~$\Model$, a prompt~$\prompt$, a possible response~$x\in\calT^\star$, and indices $i, j \in [\abs{x}]$ with $i \le j$ we define the \emph{empirical entropy on substring $[i,j]$} of~$\RModel$ responding with~$x$ to~$\prompt$ as
    \begin{align*}
        \empH^{[i,j]}(\Model,\prompt,x) := -\log \Pr\Big(&\RModel\left(\prompt\right)[i:j]=x[i:j] \ \\
        &\mid \ \RModel\left(\prompt\right)[1:(i-1)] = x[1:(i-1)]\Big).
    \end{align*}
\end{definition}
We sometimes write $H_e^{i} := H_e^{[i,i]}$ to denote the empirical entropy of a single token $i$. We remark that in expectation, Definition~\ref{def:empentropy} simply captures the entropy in the response generation. That is, we have
$$\E_{x}\left[\empH(\Model,\prompt,x)\right] = H\left(\RModel\left(\prompt\right)\right), $$
where~$x\sim \RModel\left(\prompt\right)$.

\subsection{Watermarks}
We formally define a watermarking scheme as follows.
\begin{definition}\label{def:watermark}
    A \emph{watermarking scheme} for a model $\Model$ over~$\cal{T}$ is a tuple of algorithms $\calW = (\Setup, \Wat, \Detect)$ where:
    \begin{itemize}
        \item $\Setup(1^\secpar) \to \sk$ outputs a secret key, with respect to a security parameter~$\secpar$.
        \item $\Wat_{\sk}(\prompt)$ is a randomized algorithm that takes as input a prompt $\prompt$ and generates a response in~$\cal{T}^\star$.
        \item $\Detect_{\sk}(x) \to \{\true, \false\}$ is an algorithm that takes as input a sequence~$x\in\cal{T}^\star$ outputs~$\true$ or~$\false$.
    \end{itemize}
\end{definition}

Ideally, $\Detect_{\sk}(x)$  should output $\true$ if~$x$ is generated by~$\Wat_{\sk}(\prompt)$, and should output $\false$ if~$x$ is generated independently of $\sk$. 
The former property is called \emph{completeness} and the latter \emph{soundness}.

\begin{definition}\label{def:wm_sound}
    A watermarking scheme~$\calW$ is~\emph{sound} if 
    for every security parameter $\secpar$ and token sequence~$x\in \calT^\star$ of length $\abs{x} \le \poly(\lambda)$,
    \[\Pr_{\sk \gets \Setup(1^\secpar)}[\Detect_{\sk}(x) = \true] \leq \negl.\]
\end{definition}
A scheme is sound if \emph{any} text that is generated independently from~$\sk$ has negligible probability of being detected as watermarked by~$\Detect_\sk$. Essentially, this means we will \emph{never} see a false-positive detection. 

Defining completeness requires care: It is not reasonable to require~$\Detect_\sk$ to detect any sequence~$x$ generated by~$\Wat_\sk(\prompt)$ for some $\prompt$, as it is possible that~$x$ is very short, or that~$\RModel(\prompt)$ is deterministic or has very low entropy. Instead, we require~$\Detect_\sk$ to detect watermarks only in responses for which the entropy in the generation process is high enough.

\begin{definition}\label{def:wm_complete}
    A watermarking scheme~$\calW$ is~$b(L)$-\emph{complete} if 
    for every security parameter $\lambda$ and prompt~$\prompt$ of length $\abs{\prompt} \le \poly(\lambda)$,
    \[
        \Pr_{\substack{\sk \gets \Setup(1^\lambda) \\ x \gets \Wat_\sk(\prompt)}}[\Detect_\sk(x) = \false \text{ and }  \empH\left(\Model,\prompt,x\right) \geq b\left(|x|\right)] \le \negl.
    \]
\end{definition}
Definition~\ref{def:wm_complete} guarantees that any output generated by~$\Wat_\sk$ with empirical entropy at least~$b(L)$, where~$L$ is the length of the output, will be detected as watermarked with high probability. Essentially, this means we will \emph{never} see a false-negative detection on any output of high enough empirical entropy. 
In Section~\ref{sec:nec_assump} we show that it is necessary to consider the empirical entropy of the specific output rather than the standard entropy of the entire model.

We also generalize Definition~\ref{def:wm_complete} to capture contiguous substrings of outputs.
That is, we should be able to detect a watermarked output of~$\Wat_\sk$ even if~$\Detect_\sk$ is only given a long enough contiguous substring from it.
\begin{definition}\label{def:wm_complete_ss}
    A watermarking scheme~$\calW$ is~$b(L)$-\emph{substring-complete} if 
    for every prompt~$\prompt$ and security parameter $\lambda$,
    \begin{align*}
        \Pr_{\substack{\sk \gets \Setup(1^\lambda) \\ x \gets \Wat_\sk(\prompt)}}
        \Big[\;
        \exists\; i,L \in [\abs{x}] &\text{ such that }
        \Detect_\sk(x[i:i+L]) = \false \\
        &\text{ and }  \empH^{[i:i+L]}\left(\Model,\prompt,x\right) \geq b(L)\;\Big] \le \negl.
    \end{align*}
\end{definition}

This means that every contiguous part of an output of the watermarking procedure, that has high enough empirical entropy, is detected as watermarked with high probability.
We stress that the empirical entropies in \Cref{def:wm_complete,def:wm_complete_ss} are defined with respect to the original model~$\Model$, without reference to the watermarking procedure~$\Wat_\sk$.
We also note that the empirical entropy~$\empH(\Model,\prompt,x)$ is only used as part of the definition, and is not necessarily known to $\Detect_\sk$. It is in general not possible to compute~$\empH(\Model,\prompt,x)$ without knowledge of~$\prompt$.

\subsection{Undetectable Watermarks}
Finally, we define the notion of computationally undetectable watermarking schemes.
Intuitively, a scheme is undetectable if it is infeasible to distinguish between the distributions of~$\RModel$ and~$\Wat_\sk$, even when those can be queried adaptively with arbitrary prompts. 

\begin{definition}\label{def:wm_undet}
    A watermarking scheme~$\calW = (\Setup, \Wat, \Detect)$ is \emph{undetectable} if 
    for every security parameter $\secpar$ and all polynomial-time distinguishers $D$,
    \[
        \abs{\Pr[D^{\Model,\RModel}(1^\secpar) \to 1] - \Pr_{\sk \gets \Setup(1^\secpar)}[D^{\Model,\Wat_\sk}(1^\secpar) \to 1]} \le \negl,
    \]
    where the notation $D^{\calO_1, \calO_2}$ means that $D$ is allowed to adaptively query both~$\calO_1$ and $\calO_2$ with arbitrary prompts.
\end{definition}

Note that in the above definition, we allow the distinguisher access to~$\Model$ itself as well as~$\RModel$ or~$\Wat_\sk$. The only thing that is kept secret from the distinguisher is the secret key.

It is important to remark that in any undetectable watermarking scheme, the \emph{quality of outputs} must be identical between~$\RModel$ and~$\Wat_\sk$, as otherwise it would be possible to distinguish between them.
In particular, embedding the watermark does not degrade the quality of the generated text \emph{at all}.

We finally note that a watermarking scheme can be made \emph{public} by publishing the secret key~$\sk$. Then, everyone can run the detection algorithm~$\Detect_\sk$. In particular, the scheme is no longer undetectable as~$\Detect_\sk$ can be used to distinguish between~$\RModel$ and~$\Wat_\sk$.
Nevertheless, we still maintain the property that there is no degradation in the quality of watermarked outputs, as long as the definition of ``quality'' does not depend on the secret key~$\sk$.

\subsection{Statement of our Theorems}
We are now ready to formally state the guarantees of the watermarking schemes that we present.
To warm up, in Section~\ref{sec:inefficient} we give a simple construction of an~$O\left(\lambda\right)$-complete scheme, but it only achieves a much weaker notion of soundness, and the watermarking algorithm runs in \emph{expected} $\poly(\lambda)$ time rather than strict $\poly(\lambda)$ time.
In \Cref{subsec:construct_wm}, we prove the following theorem by introducing an efficient watermarking scheme with \Cref{alg:simplified-scheme,alg:simplified-detector}.

\begin{theorem}\label{thm:wm}
    For any model~$\Model$ we construct a watermarking scheme~$\calW$ that is undetectable, sound, and~$O(\lambda \sqrt{L})$-complete. 
\end{theorem}
This means that our watermarking scheme is \emph{always} undetectable and sound, and is also complete as long as there is enough empirical entropy in the model's response.

In \Cref{sec:nec_assump} we show that it is \emph{necessary} for the completeness parameter $b(L)$ to be reasonably large, with respect to~$\lambda$.
In fact, we show that it is \emph{inherent} that low empirical entropy outputs are not watermarked in any undetectable watermarking scheme for any model.


To strengthen~\Cref{thm:wm}, we also present a modified scheme (\Cref{alg:main-scheme,alg:main-detector}) in \Cref{subsec:wm_ss} which obtains substring completeness, with similar parameters.
\begin{theorem}\label{thm:wm_ss}
    For any model~$\Model$ we construct a watermarking scheme~$\calW$ that is undetectable, sound, and~$O(\lambda \sqrt{L})$-substring-complete.
\end{theorem}

In Section~\ref{sec:remove} we discuss known attacks on methods of detecting AI-generated text. We also prove that any undetectable watermarking scheme is removable, if the model enables a fairly strong form of query access and if one is willing to expend a number of queries that grows linearly with the size of generated text. Finally, in Section~\ref{sec:open} we pose some open problems related to this work.

\section{Simplified Construction}\label{sec:inefficient}
In this section we describe a simple construction of an undetectable watermarking scheme that achieves~$\Theta\left(\lambda\right)$-completeness.
However, this scheme falls short of our main constructions in \Cref{sec:wm} in two important ways. First, it has a false-positive rate of $\varepsilon = 1/\poly(\lambda)$ instead of $\negl$. We call such a scheme~$\varepsilon$-weakly-sound. Second, the watermarking procedure~$\Wat_\sk$ is not very efficient: Its \emph{expected} run-time is polynomial in the length of the output and in~$\frac{1}{\varepsilon}$, but the worst-case running time of~$\Wat_\sk$ is unbounded.


Later, in Section~\ref{sec:wm} we present our main construction which is undetectable, sound, complete (in fact, it is even substring-complete) and efficient, yet achieves suboptimal completeness. Bridging this gap is an interesting open problem discussed in Section~\ref{sec:open}. 

Let~$b\in \mathbb{N}$ be a parameter to be chosen later; the \emph{rate of false positives} will be~$2^{-b}$.
First, let's assume that after the initialization, both~$\Wat^\calO$ and~$\Detect^\calO$ have access to a random oracle~$\calO$. This is a truly random function; that is, whenever~$\calO$ is called with a new input it returns a uniformly random string in~$\{0,1\}^b$, and it returns a consistent answer when queried with a previously queried input.

\subsection{Watermarks assuming random oracle and high min-entropy} \label{subsec:highminent}
We begin by adding another strong assumption, that the min-entropy of the response to any prompt is at least $5\secpar$.
Equivalently, let~$\prompt$ be a prompt and assume that for every~$x$, we have~$\empH(\Model,\prompt,x)\geq 5\lambda$. 
We define the watermarking scheme as follows. 
The distribution of~$\Wat^\calO(\prompt)$ is defined as the distribution of~$x\sim \RModel(\prompt)$ conditioned on~$\calO(x)=0^b$. 
Equivalently,~$\Wat^\calO(\prompt)$ repeatedly draws outputs from~$\RModel(\prompt)$ until the first time it gets a response~$x$ for which~$\calO(x)=0^b$, and then it returns~$x$.
Note that this requires~$2^b$ calls to~$\RModel(\prompt)$ in expectation.
To detect whether a string~$x$ is watermarked,~$\Detect^\calO$ simply checks whether~$\calO(x)=0^b$.
We assume that~$b\leq\lambda$ and sketch a proof for the above scheme being \emph{weakly sound}, complete and undetectable.

\paragraph{Weakly Sound:}
For any string~$x$, the value of~$\calO(x)$ is truly random by the assumption. Thus, it is detected as watermarked with probability~$2^{-b}$.

\paragraph{Complete:}
By definition,~$\Wat^\calO(\prompt)$ only produces outputs $x$ such that $\calO(x) = 0^b$, which are detected as watermarked by~$\Detect^\calO$.

\paragraph{Undetectable:}

Let~$D$ be a distinguisher that can query~$\Wat^\calO(\prompt)$ at most~$2^\lambda$ times.
In expectation, the total number of times~$\Wat^\calO(\prompt)$ queries~$\RModel(\prompt)$ to answer all queries is at most~$2^\lambda \cdot 2^b \leq 2^{2\lambda}$ (since $b \leq \secpar$ by assumption).
As the probability of each output~$x$ to be output by~$\RModel(\prompt)$ is at most~$2^{-5\lambda}$, and as~$2^{2\lambda} \ll \sqrt{2^{5\lambda}}$, with high probability~$\calO$ is never queried twice on the same input.
If~$\calO$ is never queried on the same input twice, then it simply outputs an independent random value for each query. In particular, the process of repeatedly sampling $x \sim \RModel(\prompt)$ until $\calO(x) = 0^b$ is equivalent to repeatedly sampling $x \sim \RModel(\prompt)$ until an independent, fresh random string is $0^b$ --- which is identical to simply sampling $x \sim \RModel(\prompt)$.

\subsection{Removing the high min-entropy assumption}
We next define a scheme that no longer requires the assumption about the min-entropy of the model. 
We do so by only watermarking outputs with empirical entropy higher than~$6\lambda$, corresponding to the definition of~$(6\lambda)$-complete schemes.

Let~$\prompt$ be any prompt and consider the probability 
\[
    p:=\Pr_{x\leftarrow \RModel(\prompt)}\left[\empH\left(\Model,\prompt,x\right)>6\lambda\right]
.\]
Denote by~$M_{\leq}$ the distribution~$x\sim \RModel(\prompt)$ conditioned on~$\empH\left(\Model,\prompt,x\right)\leq 6\lambda$.
Similarly, denote by~$M_>$ the distribution~$x\sim \RModel(\prompt)$ conditioned on~$\empH\left(\Model,\prompt,x\right)> 6\lambda$.
Drawing~$x\sim \RModel(\prompt)$ is equivalent to the following process: with probability~$p$, we draw a string out of the distribution~$M_>$, otherwise, we draw a string out of~$M_\le$.

Therefore, we consider the following natural algorithm for~$\Wat^\calO$.
We first flip a biased coin~$c\sim \text{Bernoulli}(p)$.
If~$c=0$ then we draw an output from the distribution~$M_\le$.
If~$c=1$, then we apply the algorithm of Section~\ref{subsec:highminent} --- that is, we draw an output from the distribution~$x\sim M_>$ conditioned on~$\calO(x)=0^b$.

This scheme is again weakly sound, and it is complete for every output $x$ with empirical entropy at least~$6\lambda$ because these outputs will always satisfy $\calO(x) = 0^b$.
If~$p\leq 2^{-\lambda}$, then undetectability is straightforward: with all but negligible probability, the distinguisher will only see outputs from~$M_\le$, which we did not change.
Otherwise,~$p>2^{-\lambda}$ and thus the distribution~$M_{>}$ is of min-entropy at least~$6\lambda - \lambda = 5\lambda$.
In particular, the construction of Section~\ref{subsec:highminent} applied to~$M_>$ is guaranteed to be undetectable.

We finally note that it is intractable to compute the value of~$p$ or the conditional distributions~$M_\le,M_>$.
We avoid their explicit computation as follows.
To implement~$\Wat^\calO$, we first draw a string~$x\sim \RModel(\prompt)$.
Computing the empirical entropy~$\empH(\Model,\prompt,x)$ given~$\Model,\prompt,x$ is straightforward.
The probability that~$\empH(\Model,\prompt,x)>6\lambda$ is of course exactly~$p$.
Thus, we can check if~$\empH(\Model,\prompt,x)\leq 6\lambda$. 
If so, we simply output~$x$.
Otherwise, we need to sample a response from~$M_>$ conditioned on its output under~$\calO$ being $0^b$. 
We can do so by repeatedly sampling~$x\sim \RModel(\prompt)$ until both~$\empH(\Model,\prompt,x)>6\lambda$ and~$\calO(x)=0^b$.
Note that the probability of success is now~$p\cdot 2^{-b}$, and thus in expectation~$\frac{1}{p} 2^b$ tries are needed, which may be very large if~$p$ is small.
On the other hand, we only reach this loop with probability~$p$; hence, the expected number of queries from~$\RModel(\prompt)$ our algorithm makes is~$1+p\cdot \frac{1}{p} 2^b = 1 + 2^b$.

\subsection{Removing the random oracle assumption}
The construction presented so far uses a random oracle~$\calO$, which is impossible to implement.\footnote{Note that we cannot sample the values of $\calO$ on-the-fly, because $\Wat$ and $\Detect$ need to agree on all of the used values.}
Often, as we also do later in Section~\ref{sec:wm}, a random oracle can be replaced with a cryptographic pseudorandom function (PRF, defined in Section~\ref{subsec:prelim}).
However, the inefficiency of~$\Wat^\calO$ requires being careful about this switch.

A PRF with a security parameter~$\lambda$ requires memory and runtime~$\poly(\lambda)$ and is guaranteed to be indistinguishable from a random oracle only to distinguishers that run in time $\poly(\lambda)$ as well.
As~$\Wat^\calO$ runs in (expected) time~$2^b$, we must choose~$b = O(\log \lambda)$ for the PRF to behave as a random oracle.
This implies that the soundness is no longer~$\negl$, but is at least~$\frac{1}{\poly(\lambda)}$.

Pseudo-code for this simplified scheme is presented in \Cref{alg:simple-scheme,alg:simple-detector}. We state the properties of this scheme in \Cref{thm:simple-thm} without proof; the proofs of these properties were sketched in the preceding two sections.

\begin{theorem} \label{thm:simple-thm}
    For any~$\lambda,\Model$ and~$b\leq O(\log \lambda)$, Algorithms~\ref{alg:simple-scheme} and~\ref{alg:simple-detector} are a watermarking scheme~$\calW$ that is undetectable,~$(6\lambda)$-complete, and~$2^{-b}$-weakly-sound.
    On expectation,~$\Wat_\sk$ makes $1+2^b$ calls to $\RModel$ to generate each response.
\end{theorem}

\begin{algorithm}
\caption{Weakly-sound watermarking algorithm $\Wat_\sk$}
\label{alg:simple-scheme}
    \KwData{A prompt ($\prompt$) and a secret key $\sk$}
    \KwResult{Watermarked text $x$}
    $x\leftarrow \RModel(\prompt)$\;
    \If{$\empH(\Model,\prompt,x)>6\lambda$}{
        \While{$F_\sk(x)\neq 0^b$ or $\empH(\Model,\prompt,x)\leq 6\lambda$}{
            $x\leftarrow \RModel(\prompt)$\;
        }
    }
    \Return{x\;}
\end{algorithm}

\begin{algorithm}
\caption{Weakly-sound watermarking detector $\Detect_\sk$}
\label{alg:simple-detector}
    \KwData{Text $x$ and a secret key $\sk$}
    \KwResult{$\true$ or $\false$}
    \uIf{$F_\sk(x)= 0^b$}
    {\Return{$\true$\;}}\Else
    {\Return{$\false$\;}}
\end{algorithm}


This construction already demonstrates the importance of our completeness definition (Definition~\ref{def:wm_complete}): Only trying to watermark outputs of high empirical entropy was crucial for this simple construction's undetectability.  As we will see in Section~\ref{sec:nec_assump}, only watermarking high empirical entropy outputs is in fact inherent to undetectable watermarking schemes.

\section{Constructing Undetectable Watermarks}\label{sec:wm}
\subsection{Reduction to a Binary Alphabet} \label{sec:binary}

For ease of presentation and analysis, we describe our watermarking scheme as operating on text encoded as a binary string. 
That is, we assume that the token set is~$\calT=\{0,1\}$.

Note that this assumption is without loss of generality: We can easily convert a model $M$ with an arbitrary token set $\calT$ into a model $M'$ with a binary token set.
First, we encode each token in $\calT$ as a distinct string in $\{0,1\}^{\log |\calT|}$; note that every codeword has length at most $\log |\calT|$. For GPT-4, the number of tokens is $\abs{\calT} = 100,277$, and thus $\log |\calT| \approx 17$ \cite{tiktoken}.
Let $E$ denote this encoding function, and let $p_i$ be a distribution over $\calT$ output by $M$. We convert $p_i$ into a series of distributions $p'_{i,j}$ for $M'$, where $p'_{i,1}$ is the distribution of the first bit of the codeword corresponding to a token sampled from $p_i$.
That is, $p'_{i,1}(0) = \Pr_{t \gets p_i}[E(t)_1]$, where $E(t)_1$ denotes the first bit of $E(t)$.
Let $b_{i,1}$ denote the bit sampled by $M'$ from $p'_{i,1}$.
Each subsequent $p'_{i,j}$ is then sampled according to the distribution of the $j^{\text{th}}$ bit of the codeword corresponding to a token sampled from $p_i$, conditioned on the previous bits being equal to $b_{i,1}, b_{i,2}, \ldots, b_{i,j-1}$.
After $M'$ samples the last bit of the current token from $p'_{i, |\calT|}$, it calls $M$ to obtain the distribution $p_{i+1}$ for the next token.

Therefore, a watermarking scheme for binary alphabets can be used on models with token alphabets of arbitrary size using the above reduction.
We note that the expected length of the encoding can be reduced by using a Huffman encoding of the token set instead of an arbitrary encoding.

\subsection{Overview of the Construction}\label{subsec:overview}
In \Cref{sec:inefficient}, we saw a simple scheme that plants a watermark by sampling only texts for which an easily checkable predicate holds.
In order to make this scheme more efficient, a natural idea is to sample tokens one at a time.
If we don't require undetectability, an easy way to do this is to use a $\{0,1\}$-valued hash function $h$ and sample tokens $x_j$ with preference for those satisfying $h(x_j) = 1$.
Given some text, we can determine whether a watermark is present by computing the hash of each token.
In watermarked text, more tokens should hash to 1 than to 0; in un-watermarked text, there should be no bias. 
This is a classic idea in steganography, discussed in \cite{steg}. It is essentially the idea used in~\cite{kgw}.

Unfortunately, this strategy significantly alters the output distribution, making it easily detectable: It prefers half of the words in the token set. 
As long as a biasing strategy yields a significant expected gap between the incidence of some predicate in watermarked text versus natural text, the resulting scheme should yield an observable watermark.
Our objective is to plant a signal \emph{without} noticeably changing the distribution of each token.


We first discuss a watermarking scheme that can only be used to generate a single output text of a predetermined maximum length $L$, for an arbitrary prompt.
The secret key shared by the watermarked model and the detection algorithm will be a sequence~$\vec u = u_1,\dots,u_L$ of uniformly chosen real numbers in the range~$[0,1]$.
Even though this state is independent of the prompt the model will receive, we show that this shared state is enough to plant a watermark in any single response. From the perspective of a user who doesn't know the secret key~$\vec u$, the distribution of outputs is not changed at all.

When generating a response, the watermarked model will use the secret key to decide on each output token. 
Consider the generation of the~$j$-th token in the response, after the previous tokens are already decided.
Let $p_j(1)$ denote the probability, according to the real model, of this token being~$1$.
The watermarked model outputs~$x_j=1$ if $u_j \leq p_j(1)$ and~$x_j=0$ otherwise.
As~$u_j$ was drawn uniformly from~$[0,1]$, the probability that the watermarked model output~$x_j=1$ is \emph{exactly} $p_j(1)$. Therefore, the distribution of generated text (in a single response) does not change at all.
Nevertheless, we next show that the detection algorithm can compare the generated text to the shared sequence $\vec u$, and deduce that the generated output was drawn from the watermarked model.

For each text bit~$x_j$, the detection algorithm can compute a score
\[
    s(x_j,u_j) = \begin{cases}
             \ln \frac{1}{u_j}  & \text{if } x_j = 1 \\
             \ln \frac{1}{1-u_j} & \text{if } x_j = 0
       \end{cases} .
\]
Given a string~$x=(x_1,\ldots,x_L)$, the detection algorithm sums the score of all text bits 
\[
    c(x) = \sum_{j=1}^L s(x_j,u_j).
\]
Crucially, the detection algorithm does not need to know the distributions with which the model produces output bits. Since the detection algorithm does not have access to the prompt, it would not be able to compute those distributions.

We observe that the expected score is higher in watermarked text, as~$u_j$ is correlated with the output bit~$x_j$.
In non-watermarked text, the value of~$u_j$ is independent of the value of~$x_j$.
Therefore,~$s(x_j,u_j)$ is simply an exponential random variable with mean~$1$:
\[
    \E_{u_j}[s(x_j,u_j)] = \int_0^1 \ln(1/x) \ dx = 1,
\]
and we have $\E_{\vec u}[c(x) - \abs{x}] = 0$.

For watermarked outputs, on the other hand,
	\begin{align*}
		\E_{u_j}[s(x_j,u_j)] &= \int_0^{p_j(1)} \ln \frac{1}{u} \ du + \int_{p_j(1)}^1 \ln \frac{1}{1-u} \ du \\
		&= \int_0^{p_j(1)} \ln \frac{1}{u} \ du + \int_0^{p_j(0)} \ln \frac{1}{u} \ du \\
		&= \big(p_j(1) - p_j(1) \cdot \ln p_j(1)\big) + \big(p_j(0) - p_j(0) \cdot \ln p_j(0)\big) \\
		&= 1 + \ln(2) \cdot H(p_j),
	\end{align*}
and the total expected score is
\[
    \E_{\vec u}[c(x) - \abs{x}] = \ln 2 \cdot H(\RModel(\prompt)).
\]

We've shown that there's a substantial gap between the expected scores of watermarked and natural text, as long as the text generation has high entropy.
This should give us hope that this biasing strategy yields a reliable detector, but there are a few obstacles left on the way.

First, the expectation argument turns out to not be very useful because the variance of the score could be large.
In Section~\ref{sec:nec_assump} we discuss why this implies that we must consider \emph{empirical entropy} instead of the entropy of the entire model. In Sections~\ref{subsec:construct_wm} and~\ref{subsec:wm_ss} we use empirical entropy to build effective distinguishers.

Second, the scheme described above is only indistinguishable for a single response, and that response must be shorter than the secret key. 
A natural idea is to use a psuedorandom function (refer to \Cref{subsec:prelim} for a definition) to determine the values~$u_j$, instead of drawing them all in advance.
For example, by setting~$u_j= F_\sk(j)$ the length of any single response no longer has to be bounded. 
As $F_\sk$ is queried on each input~$j$ at most once, the values of $u_j$ are pseudorandom and the distribution of a single watermarked output is computationally indistinguishable from the original distribution.
The question becomes: Can we deal with multiple responses?
One of our main contributions, and the most substantial difference from all prior work, is to answer this question in the affirmative.

Let~$r^{(i)}$ be a unique identifier assigned to each response.
This might be a global counter or a random string (usually referred to as a \emph{nonce}).
To sample the $j$-th token of the $i$-th response we can use $u_j^{(i)} = F_\sk(r^{(i)},j)$.
If all pairs $(r^{(i)},j)$ are unique, then the values of $u^{(i)}_j$ are pseudorandom.
However, the detection algorithm needs to know~$r^{(i)}$ to compute the detection score.
If~$r^{(i)}$ is a counter, then we would need to keep a global state to maintain it. Moreover, to use the detection algorithm we would need to enumerate over all possible counter values.
If~$r^{(i)}$ is a long random string, no global state is needed, but the detection algorithm still needs to know~$r^{(i)}$.
While~$r^{(i)}$ must be recoverable by the detection algorithm, it cannot simply be written in the output text, as we might as well just append ``WATERMARK!'' to it instead (which would obviously change the distribution of outputs).
Our solution is to use real randomness to generate the first few tokens of each output, keeping track of how much \emph{entropy} we used in the process. Once this entropy passes some specified threshold, we use the high-entropy prefix as $r^{(i)}$. Since these prefixes have high enough entropy, all choices of~$r^{(i)}$ will be unique with all but negligible probability. The detection algorithm will test whether any prefix in the text, if used as~$r^{(i)}$, will yield an unusually high score for the remainder of the text.
The details of this construction are presented in \Cref{subsec:construct_wm}.

In the above sketch the detector needs the entire output from the model to detect the watermark. 
We describe a modification of this scheme in \Cref{subsec:wm_ss} which is able to detect the watermark, even when it is given only an contiguous substring of the output with sufficiently high entropy. 
Essentially, this modification works the same except it ``resets'' the choice of~$r^{(i)}$ whenever enough \emph{new} entropy is observed.

\subsection{Constructing Undetectable Watermarks}\label{subsec:construct_wm}
Let $\poly_1(\cdot),\poly_2(\cdot)$ be polynomials.   
Let $F_\sk : \{0,1\}^{\poly_1(\secpar)} \to \{0,1\}^{\poly_2(\secpar)}$ be a PRF, where $\sk \in \{0,1\}^\secpar$. We wish to interpret the output of $F_\sk$ as a real number in $[0,1]$. We do so by letting $z$ be the integer representation of the output and taking $\frac{z}{2^{\poly_2(\secpar)}}$. We consider $\poly_2$ to be a large polynomial and ignore floating point errors.
In the below algorithms, we allow $F_\sk$ to take strings of varying length as input; we assume that $\poly_1(\cdot)$ is chosen such that these strings are never too long, and if they are too short we pad them. 
In this section we assume that the token alphabet is binary as discussed in \Cref{sec:binary}.
We let $\texttt{done}$ denote the binary encoding of the ``done" token, and we write $\texttt{done} \in (x_1, \ldots, x_k)$ if and only if the decoding of $(x_1, \ldots, x_k)$ in the original token alphabet includes $\texttt{done}$.

\begin{algorithm}
\caption{Complete watermarking algorithm $\Wat_\sk$}
\label{alg:simplified-scheme}
    \KwData{A prompt ($\prompt$) and a secret key $\sk$}
    \KwResult{Watermarked text $x_1, \hdots, x_L$}
    $i \gets 1$\;
    $H \gets 0$\;
    \While{$\texttt{done} \notin (x_1, \ldots, x_{i-1})$}{
        $p_i \gets \Model(\prompt, x_1, \hdots, x_{i-1})$\;
        \uIf{$H < \lambda$}{
            \tcp{Collect more internal entropy}
            Sample $x_i \gets p_i$\;
            $H \gets H - \log p_i(x_i)$\;
            \If{$H \ge \lambda$}{
                $r \gets (x_1, \dots, x_i)$\; \label{line:simplified-prefix}
            }
        }
        \Else{
            \tcp{Embed the watermark}
            $x_i \gets \mathbbm{1}[F_\sk(r,i) \le p_i(1)]$\;
        }
        $i \gets i + 1$\;
        }
\end{algorithm}

\begin{algorithm}
\caption{Complete detector $\Detect_\sk$}
\label{alg:simplified-detector}
    \KwData{Text $x_1, \ldots, x_L$ and a secret key $\sk$}
    \KwResult{$\true$ or $\false$}
    \For{$i \in [L]$}{
        $r^{(i)} \gets (x_1, \hdots, x_{i})$\;
        Define $v^{(i)}_j := x_j \cdot F_\sk(r^{(i)},j) + (1-x_j) \cdot (1-F_\sk(r^{(i)},j))$ for $j \in [L]$\;
        \If{$\sum_{j=i+1}^L \ln(1/v^{(i)}_j) > (L-i) + \lambda \sqrt{L-i}$}{
        	\Return $\true$\;
        }
    }
    \Return $\false$\;
\end{algorithm}

In this section we let $\calW = (\Setup, \Wat, \Detect)$ denote the watermarking scheme where $\Wat$ is \Cref{alg:simplified-scheme}, $\Detect$ is \Cref{alg:simplified-detector}, and $\Setup(1^\secpar)$ samples $\sk \gets \{0,1\}^\secpar$. This scheme is outlined above in \Cref{subsec:overview}.

Let $\Wat^\calO$ and $\Detect^\calO$ be the same algorithms as $\Wat_\sk$ and $\Detect_\sk$, except that they use a random oracle $\calO$ instead of $F_\sk$. Since both algorithms only make black-box use of $F_\sk$, these are well-defined. Denote this random oracle scheme by $\calW^\calO$ (which does not need a $\Setup$ algorithm). Undetectability, $b(L)$-(substring-)completeness, and soundness are defined identically for $\calW^\calO$, except we replace the probabilities over $\sk \from \Setup(1^\secpar)$ with probabilities over $\calO \from \{f : \{0,1\}^{\poly_1(\secpar)} \to \{0,1\}^{\poly_2(\secpar)}\}$. Note that in the definition of undetectability for $\calW^\calO$, the distinguisher will not be given access to the random oracle $\calO$ (since the distinguisher for $\calW$ is not given access to $F_\sk$).

\begin{lemma}
\label{lemma:prf-switch}
    The watermarking scheme $\calW$ is undetectable/$b(L)$-(substring-)complete/sound if and only if  $\calW^\calO$ is undetectable/$b(L)$-(substring-)complete/sound, assuming the security of the PRF used in $\calW$.
\end{lemma}
\begin{proof}
    The security of the PRF says that black-box access to $F_\sk$ (for random $\sk$) is indistinguishable from black-box access to a random $\calO$, for any polynomial-time distinguisher. Observe that \Cref{alg:simplified-scheme,alg:simplified-detector} both only make black-box use of $F_\sk$. Therefore it is possible to efficiently test, using only black-box access to $F_\sk$, whether a given text/prompt/distinguisher violates soundness/$b(L)$-(substring-)completeness/undetectability. The security of the PRF then implies that the advantage of any given text/prompt/distinguisher is at most $\negl$ different between $\calW$ and $\calW^\calO$.
\end{proof}

\begin{lemma} \label{lemma:exponential-concentration}
    Let $E_1, \dots, E_n$ be i.i.d. exponential random variables with rate 1, and let $\overline{E} := \sum_{i=1}^n E_i$ be their sum. Then for any $\tau > 0$,
    \begin{itemize}
        \item[(a)] $\Pr[\overline{E} \ge n + \sqrt{\tau n}] \le \left(\frac{4}{5}\right)^{\sqrt{\tau}}$, and
        \item[(b)] $\Pr[\overline{E} \le n - \sqrt{\tau n}] \le e^{-\tau/2}$.
    \end{itemize}
\end{lemma}
\begin{proof}
    We start with part (a). By \cite[Theorem 5.1(i)]{Jan18},
    \begin{align*}
        \Pr[\overline{E} \ge n + \sqrt{\tau n}] &\le e^{-n \cdot \left(\sqrt{\tau/n} - \ln(1+\sqrt{\tau/n})\right)} \\
        &= \left(\frac{e^{\sqrt{\tau/n}}}{1+\sqrt{\tau/n}}\right)^{-n}.
    \end{align*}
    If $\tau \ge n$, then since $1+z \le 2^z$ for $z \ge 1$ we have
    \begin{align*}
        \left(\frac{e^{\sqrt{\tau/n}}}{1+\sqrt{\tau/n}}\right)^{-n} &\le \left(\frac{e}{2}\right)^{-\sqrt{\tau n}} \\
        &\le \left(\frac{4}{5}\right)^{\sqrt{\tau n}}.
    \end{align*}
    If $\tau \le n$, then using $e^z \ge 1 + z + z^2/2$ for $z \ge 0$ we have
    \begin{align*}
        \left(\frac{e^{\sqrt{\tau/n}}}{1+\sqrt{\tau/n}}\right)^{-n} &\le \left(\frac{1+\sqrt{\tau/n} + \tau/2n}{1+\sqrt{\tau/n}}\right)^{-n} \\
        &= \left(1 + \frac{\tau/2n}{1+\sqrt{\tau/n}}\right)^{-n} \\
        &\le \left(1 + \frac{\tau}{4n}\right)^{-n} \\
        &\le \left(1 + \frac{1}{4}\right)^{-\tau} \\
        &= \left(\frac{4}{5}\right)^{\tau},
    \end{align*}
    where we have also used the fact that $(1+\frac{z}{n})^n$ is monotonically increasing in $n$.

    We now turn to part (b). By \cite[Theorem 5.1(iii)]{Jan18},
    \begin{align*}
        \Pr[\overline{E} \le n - \sqrt{\tau n}] &\le e^{n \cdot \left(\sqrt{\tau/n} + \ln(1-\sqrt{\tau/n})\right)} \\
        &= e^{\sqrt{\tau n}} \cdot \left(1-\sqrt{\tau/n}\right)^n.
    \end{align*}
    If $\tau \ge n$, the probability becomes 0. If $\tau < n$, then taking the natural logarithm the above becomes
    \begin{align*}
        \sqrt{\tau n} + n \ln \left(1-\sqrt{\tau/n}\right) &= \sqrt{\tau n} \cdot \left(1+\frac{\ln(1-\sqrt{\tau/n})}{\sqrt{\tau/n}}\right) \\
        &\le \sqrt{\tau n} \cdot \left(1-\frac{2}{2-\sqrt{\tau/n}}\right) \\
        &\le -\tau/2
    \end{align*}
    where for $0 < z < 1$ we have used the facts that $\frac{\ln(1-z)}{z} \le \frac{-2}{2-z}$ and $\frac{2}{2-z} \ge 1 + \frac{z}{2}$.
\end{proof}

\begin{theorem} \label{thm:wm_sound}
    $\calW$ is a sound watermarking scheme.
\end{theorem}
\begin{proof}
    Recall the definition of soundness in \Cref{def:wm_sound}. By \Cref{lemma:prf-switch} it suffices to show that for any text $x = x_1, \dots, x_L$,
    \[
        \Pr_{\calO}[\Detect^\calO(x) = \true] \le \negl.
    \]
    For $i,j \in [L]$, define $r^{(i)}$, $v_j^{(i)}$ as in \Cref{alg:simplified-detector} and let $u_j^{(i)} := \calO(r^{(i)},j)$. Recall that $v_j^{(i)} = x_j \cdot u_j^{(i)} + (1-x_j) \cdot (1-u_j^{(i)})$.
    
    Since $u_j^{(i)}$ is independent from $x_j$, we have $v_j^{(i)} \sim U([0,1])$. Therefore, $E_j^{(i)} := \ln(1/v_j^{(i)})$ are independent exponential random variables with rate parameter 1. By \Cref{lemma:exponential-concentration},
    \[
        \Pr[\sum_{j=i+1}^L E_j^{(i)} > (L-i) + \lambda \sqrt{L-i}] \le \left(\frac{4}{5}\right)^{\lambda}.
    \]
    By a union bound over all $L$ possible values of $i$, the probability of \Cref{alg:simplified-detector} returning $\true$ is at most $L \cdot (4/5)^{\lambda} = \negl$, completing the proof.
\end{proof}

\begin{theorem}\label{thm:wm_complete}
    $\calW$ is a $\left(\frac{4}{\ln 2} \lambda \sqrt{L}\right)$-complete watermarking scheme.
\end{theorem}
\begin{proof}
    Recall the definition of completeness in \Cref{def:wm_complete}. By~\Cref{lemma:prf-switch} it suffices to show that for every~$\prompt$,
    \begin{equation}
        \Pr_{\substack{\calO \\ x \gets \Wat^\calO(\prompt)}}[\Detect^\calO(x) = \false \text{ and }  \empH\left(\Model, \prompt, x\right) \geq b(\abs{x})] \le \negl \label{eq:oracle-completeness}
    \end{equation}
    where $b(L) = \frac{4}{\ln 2} \lambda \sqrt{L}$.
    In fact, we will prove something stronger: For \emph{every} fixed~$x\in \calT^\star$ and~$\prompt$ such that~$\empH(\Model, \prompt, x) \geq \frac{4}{\ln 2} \lambda \sqrt{\abs{x}}$, if each bit $x_i$ of $x$ has empirical entropy $H_e^{i}(\Model,\prompt,x) \le \lambda$,
    \begin{equation}
        \Pr_{\calO}\left[\Detect^\calO(x) = \false \;\mid\; \Wat^\calO(\prompt)=x\right] \le \negl. \label{eq:oracle-conditional-completeness}
    \end{equation}
    Inequality \ref{eq:oracle-conditional-completeness} says that for any possible fixed output~$x$ that has high empirical entropy (which isn't too concentrated on any particular bit), conditioning on~$\Wat^\calO$ outputting it, it is likely to be detected as watermarked. Note that the probability here is over the choice of outputs of $\calO$ and not over~$x$, which is fixed.

    Observe that Inequality~\ref{eq:oracle-conditional-completeness} is not falsifiable, so we cannot do the PRF switch (\Cref{lemma:prf-switch}) with it. However, since each bit has at most a $2^{-\lambda}$ chance of having empirical entropy more than $\lambda$, Inequality~\ref{eq:oracle-conditional-completeness} implies Inequality~\ref{eq:oracle-completeness} via the law of total probability.
    
    In order to prove Inequality \ref{eq:oracle-conditional-completeness}, we just need to show that the correct choice of prefix $r$, determined on \Cref{line:simplified-prefix} of \Cref{alg:simplified-scheme}, has a high score (since $\Detect$ tries every possible prefix). Let $x = x_1, \dots, x_L$ where $L := \abs{x}$, and let $\ell := \abs{r}$ where $r$ is the correct prefix.
    For $j\in [\ell+1,L]$, let $v_j := x_j \cdot u_j + (1-x_j) \cdot (1-u_j)$.
    We prove that~$\sum_{j=\ell+1}^L \ln \frac{1}{v_j}$ is likely to be larger than the detection threshold~$(L-\ell) + \lambda \sqrt{L-\ell}$.
    
    Denote by~$s_j$ the random variable~$\ln \frac{1}{v_j}$, conditioned on~$\Wat^\calO(\prompt)=x$, or equivalently, on the value of~$x_j$.
    Recall that the variable~$E=\ln \frac{1}{u}$ for~$u\sim U([0,1])$ is exponentially distributed with rate~$1$.
    In particular, if~$x_j=1$ the variable~$s_j$ is distributed the same as~$E$ conditioned on~$u \leq p_j(1)$, or equivalently on~$E\geq \ln \frac{1}{p_j(1)}$.
    By the memorylessness property of exponential distributions, hence, $s_j$ is distributed as~$\ln \frac{1}{p_j(1)} + E_j$, where~$E_j$ is an exponentially distributed random variable with rate~$1$.
    Symmetrically, if~$x_j=0$ the variable~$s_j$ is distributed as~$\ln \frac{1}{p_j(0)} + E_j$.
    We conclude that
    \[
        \sum_{j=\ell+1}^L s_j \sim \ln(2) \cdot \empH^{[\ell+1,L]}(\Model,\prompt,x) + \sum_{j=\ell+1}^{L} E_j,
    \]
    where the factor of $\ln(2)$ comes from the switch to binary entropy.
    Recall that $\Wat_\sk$ fixes $r$ as soon as it outputs $x_\ell$ for which $(x_1, \ldots, x_\ell)$ has empirical entropy of at least $\secpar$.
    Since each bit, including $x_\ell$, has empirical entropy of at most $\secpar$, $(x_1, \ldots, x_\ell)$ has empirical entropy at most $2\secpar \le \frac{2}{\ln 2} \lambda \sqrt{L}$.
    Therefore,
    \begin{align*}
    \empH^{[\ell+1,L]}(\Model,\prompt,x) &\geq 
    \empH(\Model,\prompt,x) - 2\lambda \\
    &\geq \frac{2}{\ln 2} \lambda \sqrt{L} \\
    &\geq \frac{2}{\ln 2} \lambda \sqrt{L-\ell}.
    \end{align*}
    Therefore, applying \Cref{lemma:exponential-concentration},
    \begin{align*}
        \Pr[\sum_{j=\ell+1}^L s_\ell \le (L-\ell) + \lambda \sqrt{L-\ell}]
        &\leq \Pr[\sum_{j=\ell+1}^{L} E_j \le (L-\ell) - \lambda \sqrt{L-\ell}] \\
        &\le e^{-\lambda^2/2}. \qedhere
    \end{align*}
\end{proof}


\begin{theorem} \label{thm:wm_undet}
    $\calW$ is an undetectable watermarking scheme.
\end{theorem}
\begin{proof}
    Recall \Cref{def:wm_undet}, which says that a watermark is undetectable if no efficient adversary can distinguish between query access to the watermarked model and the original one. Consider any fixed history of responses $x^{(1)}, \dots, x^{(t-1)}$, and suppose that the adversary submits $\prompt$ as the next query. We will show that
    \[
        \frac{1}{2} \norm{\Wat^\calO(\prompt) - \RModel(\prompt)}_1 \le \negl.
    \]
    Since the adversary can only make $\poly(\lambda)$ queries, it follows that the entire interaction with $\Wat^\calO$ is statistically indistinguishable from interaction with $\RModel$. Finally, we will obtain the theorem by invoking \Cref{lemma:prf-switch}.
    
    Let $r^{(1)}, \dots, r^{(t-1)}$ be the prefixes of the previous responses $x^{(1)}, \dots, x^{(t-1)}$ (as determined on \Cref{line:simplified-prefix} of \Cref{alg:simplified-scheme}). If for some $k \in [t]$ the watermarking scheme never collects enough entropy to assign a prefix $r^{(k)}$, we let $r^{(k)} := \bot$. We denote the tokens of $x^{(k)}$ by $(x^{(k)}_1, \dots, x^{(k)}_{L^{(k)}})$ where $L^{(k)} := \abs{x^{(k)}}$, and the corresponding probability distributions output by $\Model$ with $(p^{(k)}_1, \dots, p^{(k)}_{L^{(k)}})$. We denote the tokens of $r^{(k)}$ by $(r^{(k)}_1, \dots, r^{(k)}_{\ell^{(k)}})$ where $\ell^{(k)} := \abs{r^{(k)}}$.
    
    Since we have fixed $x^{(1)}, \dots, x^{(t-1)}$, observe that the distribution on the next prefix $r^{(t)}$ is identical between $\Wat^\calO$ and $\RModel$. This is because $\Wat^\calO$ does not start embedding the watermark until after $r^{(t)}$ is completely sampled; until then $\Wat^\calO$ samples tokens according to $\RModel$. Define the set $B := \{r^{(1)}, \dots, r^{(t-1)}\} \setminus \{\bot\}$. For any fixed $r^{(t)} \not\in B$, the distribution on the remaining tokens $(x^{(t)}_{\ell^{(t)}+1}, \dots, x^{(t)}_{L^{(t)}})$ is also identical between $\Wat^\calO$ and $\RModel$: If $r^{(t)} = \bot$, then there are no remaining tokens and the statement is trivial; if $r^{(t)} \not\in \{r^{(1)}, \dots, r^{(t-1)}\}$ then $\Wat^\calO$ samples the remaining tokens with fresh randomness. We will show that $\Pr_{r^{(t)}}[r^{(t)} \in B] \le \negl$, completing the proof.
    
    For $k \in [t-1]$ and $i \in [L^{(k)}]$, we define
    \[
        q^{(k)}_i := \Model(\prompt, x^{(k)}_1, \dots, x^{(k)}_{i-1}).
    \]
    Note that $q^{(k)}_i(z)$ is the probability that $x^{(t)}_i = z$, given that $x^{(t)}_j = x^{(k)}_j$ for $j \in [i-1]$. We compute
    \begin{align*}
        \Pr_{r^{(t)}}[r^{(t)} \in B] &= \Pr_{r^{(t)}}[r^{(t)} \in \{r^{(1)}, \dots, r^{(t-1)}\} \text{ and } r^{(t)} \ne x^{(t)}] \\
        &\le \sum_{k=1}^{t-1} \Pr_{r^{(t)}}[r^{(t)} = r^{(k)} \text{ and } r^{(t)} \ne \bot] \\
        &= \sum_{k=1}^{t-1} \mathbbm{1}\left[\sum_{i=1}^{\ell^{(k)}-1} \log \frac{1}{q_i^{(k)}(x_i^{(k)})} < \lambda \le \sum_{i=1}^{\ell^{(k)}} \log \frac{1}{q_i^{(k)}(x_i^{(k)})}\right] \cdot \prod_{i=1}^{\ell^{(k)}} q_i^{(k)}(r_i^{(k)}) \\
        &\le \sum_{k=1}^{t-1} \mathbbm{1}\left[\lambda \le -\log \prod_{i=1}^{\ell^{(k)}} q_i^{(k)}(r_i^{(k)})\right] \cdot \prod_{i=1}^{\ell^{(k)}} q_i^{(k)}(r_i^{(k)}) \\
        &\le (t-1) \cdot 2^{-\lambda}. \qedhere
    \end{align*}
\end{proof}


\subsection{Constructing Substring-Complete Watermarks} \label{subsec:wm_ss}
The detector presented in~\Cref{subsec:construct_wm} receieves as an input the entire text output by $\Wat_\sk$. In this section we generalize the scheme into a substring-complete one, which is able to detect watermarks in any contiguous sequence of text with sufficiently high empirical entropy.

The new scheme, described in \Cref{alg:main-scheme,alg:main-detector}, is essentially a repeated version of \Cref{alg:simplified-scheme,alg:simplified-detector}. First, it samples naturally from the model until it has collected enough empirical entropy. Once we collect $\lambda$ bits of empirical entropy in a text block $r$, we start embedding the watermark using $r$ as our seed. 
We continue embedding the watermark in the next subsequence of tokens $(x_i, \ldots, x_{i+\ell-1})$ until we have collected enough empirical entropy to know that the watermark will be detected with high probability.
At this point, we could restart \Cref{alg:simplified-scheme}, sampling the next tokens from $x_j \gets p_j$ to generate a new seed $r$.
Instead, we observe that we can in fact use the previous subsequence $(x_i, \ldots, x_{i+\ell-1})$ itself as $r$, as it is of high enough empirical entropy as well.

\begin{algorithm}
\caption{Substring-complete watermarking algorithm $\Wat_\sk$.}
\label{alg:main-scheme}
    \KwData{A prompt ($\prompt$), secret key $\sk$, and parameter $\secpar$}
    \KwResult{Watermarked text $x_1, \hdots, x_L$}
    $r \gets \bot, H \gets 0, i \gets 1, \ell \gets 0$\;
    \While{$\texttt{done} \notin (x_1, \ldots, x_{i-1})$}{
        $p_i \gets \Model(\prompt, x_1, \hdots, x_{i-1})$\;
        \uIf{$r = \bot$}{
            \tcp{Still sampling the first block}
            Sample $x_i \from p_i$\;
        } \uElse{
            \tcp{Embed the watermark}
            $x_i \gets \mathbbm{1}[F_\sk(r,i) \le p_i(1)]$\;
        }
        $H \gets H - \log p_i(x_i)$\;
        \If{$H \ge \frac{2}{\ln 2} \lambda \sqrt{\ell}$}{
            \tcp{Reassign $r$}
            $r \gets (x_{i-\ell}, \dots, x_i)$\; \label{line:assign-randomness}
            $H \gets 0, \ell \gets 0$\;
        }
        $i \gets i + 1, \ell \gets \ell+1$\;
    }
\end{algorithm}

\begin{algorithm}
\caption{Substring-complete detector $\Detect_\sk$.}
\label{alg:main-detector}
    \KwData{Text $x_1, \ldots, x_L$ and a secret key $\sk$}
    \KwResult{$\true$ or $\false$}
    \For{$i,\ell \in [L]$, $\ell < i$}{
        $r^{(i,\ell)} \gets (x_{i-\ell}, \hdots, x_i)$\;
        $v_j^{(i,\ell)} \gets x_j \cdot F_\sk(r^{(i,\ell)},j) + (1-x_j) \cdot (1-F_\sk(r^{(i,\ell)},j))$ for each $j \in [L]$\;
        \For{$k \in [i+1,L]$}{
            \If{$\sum_{j=i+1}^k \ln(1/v_j^{(i,\ell)}) > (k-i) + \lambda \sqrt{k-i}$}{
            	\Return $\true$\;
            }
        }
    }
    \Return $\false$\;
\end{algorithm}

We use the same notation as in \Cref{subsec:construct_wm}, except that now $\calW$ refers to the scheme defined in \Cref{alg:main-scheme,alg:main-detector}.

\begin{theorem}
    $\calW$ is a sound watermarking scheme.
\end{theorem}
\begin{proof}
    Up to symbolic differences, the proof is identical to that of \Cref{thm:wm_sound} except that the union bound will be over $L^3$ possible values of $i, \ell, k$ rather than $L$ possible values of $i$.
\end{proof}
 
\begin{theorem}
    $\calW$ is a $\left(\frac{8}{\ln 2} \lambda \sqrt{L}\right)$-substring-complete watermarking scheme.
\end{theorem}

\begin{proof}
    We argue that any substring with empirical entropy at least $\frac{8}{\ln 2} \lambda \sqrt{L}$ must include the entirety of a pair of consecutive blocks $r^{(a)}, r^{(a+1)}$.
    We then refer to the proof of \Cref{thm:wm_complete} to argue that the bias planted in $r^{(a+1)}$ with the seed $r^{(a)}$ can be detected with overwhelming probability.
    
    Recall the definition of completeness in \Cref{def:wm_complete}. By~\Cref{lemma:prf-switch} it suffices to show that for every~$\prompt$,
    \begin{align}
        \Pr_{\substack{\calO \\ x \gets \Wat^\calO(\prompt)}}
        \Big[\;
        \exists\; i,L \in [\abs{x}] &\text{ such that }
        \Detect_\sk(x[i:i+L]) = \false \nonumber \\
        &\text{ and }  \empH^{[i:i+L]}\left(\Model,\prompt,x\right) \geq b(L)\;\Big] \le \negl.
    \end{align}
    where $b(L) = \frac{8}{\ln 2} \lambda \sqrt{L}$.
    We'll show that for \emph{every} fixed~$x\in \calT^\star$, $i, L \in [|x|]$, and~$\prompt$ such that~$\empH^{[i:i+L]}(\Model, \prompt, x) \geq \frac{8}{\ln 2} \lambda \sqrt{L}$, if each bit $x_j$ of $x[i:i+L]$ has empirical entropy $H_e^{j}(\Model,\prompt,x) \le \lambda$,
    \begin{equation}
        \Pr_{\calO}\left[\Detect^\calO(x) = \false \;\mid\; \Wat^\calO(\prompt)=x\right] \le \negl. 
    \end{equation}
    Let $x = (x_i, \ldots x_{i+L})$ be such that $\empH^{[i:i+L]}(\Model, \prompt, x) \geq \frac{8}{\ln 2} \lambda \sqrt{L}$ and $H_e^{j}(\Model,\prompt,x) \le \lambda$ for each $j \in [i:i+L]$. 
    Let $c,d \in [i:i+L]$ be such that $r^{(a)} := (x_{c}, \ldots, x_{d})$ is the first contiguous substring of $x[i:i+L]$ such that $r$ was assigned to $r^{(a)}$ in \Cref{line:assign-randomness} of \Cref{alg:main-scheme}.
    Recall that we reassign each seed $r$ as soon as we have collected $\frac{2}{\ln 2} \lambda \sqrt{\ell}$ empirical entropy where $\ell \leq L$.
    So $x[i: c - 1]$ contains at most $\frac{2}{\ln 2} \lambda \sqrt{L} + \lambda$ empirical entropy; otherwise, $r^{(a)}$ would have been assigned for a lower starting index.
    Similarly, $r^{(a)}$ contains at most $\frac{2}{\ln 2} \lambda \sqrt{L} + \secpar$ empirical entropy.
    Let $r^{(a+1)} = (x_{d+1}, \ldots, x_k)$ denote the next block after $r^{(a)}$.
    Observe that 
    $$\empH^{[d+1:i+L]}(\Model, \prompt, x) \geq \frac{8}{\ln 2} \lambda \sqrt{L} - \frac{4}{\ln 2} \lambda \sqrt{L} - 2\secpar \geq \frac{2}{\ln 2} \lambda \sqrt{k-d}.$$

    Therefore, after $r^{(a)}$ has been fixed, there is sufficient entropy in the remainder of our substring $x[d+1:i+L]$ so that $r^{(a+1)}$ will be contained entirely within it.
    Since $r^{(a+1)}$ has empirical entropy at least $\frac{2}{\ln 2} \lambda \sqrt{k-d}$, we can now follow the analysis in the proof of \Cref{thm:wm_complete} to conclude that this choice of $c,d,k$ in the detection algorithm results in an output of $\true$.
\end{proof}

\begin{theorem}
    $\calW$ is an undetectable watermarking scheme.
\end{theorem}
\begin{proof}
    The proof is similar to \Cref{thm:wm_undet}, but we index by ``blocks'' instead of queries. These serve the same purpose as the ``prefixes'' of \Cref{thm:wm_undet}, except that there may be many blocks in any single query. We define a \emph{block} to be a sequence of tokens $x_{i-\ell}, \dots, x_i$ defining a value of $r$ on \Cref{line:assign-randomness} of \Cref{alg:main-scheme}. If the last block in a response does not get completed before reaching a $\done$ token, we assign that block the value $\bot$. Over the course of the distinguishing experiment that defines undetectability in \Cref{def:wm_undet}, $\Wat^\calO$ will output some number of blocks in each response. We enumerate all blocks $r^{(1)}, r^{(2)}, \dots$ that $\Wat^\calO$ outputs during the course of the experiment (across all responses).
    
    Recall \Cref{def:wm_undet}, which says that a watermark is undetectable if no efficient adversary can distinguish between query access to the watermarked model and the original one. Let $\calR_W$ and $\calR_M$ be the distributions over the next block $r^{(t)}$ under $\Wat^\calO$ and $\RModel$, respectively.

    We will prove the following:
    \begin{enumerate}
        \item[] For any fixed history of blocks $r^{(1)}, \dots, r^{(t-1)}$, if
        \[
            r^{(t-1)} \not\in \{r^{(1)}, \dots, r^{(t-2)}\} \setminus \{\bot\}
        \]
        then $\calR_W = \calR_M$ and
        \begin{align}
        \Pr_{r^{(t)}}\left[r^{(t)} \in \{r^{(1)}, \dots, r^{(t-1)}\} \setminus \{\bot\}\right] &\le \negl. \label{ineq:substring-good}
    \end{align}
    \end{enumerate}
    Inductively, any $\poly(\lambda)$-many blocks output by $\Wat^\calO$ are at most $\negl$-far in statistical distance from outputs of $\RModel$. Since only $\poly(\lambda)$ blocks can appear in the experiment, the entire interaction with $\Wat^\calO$ is statistically indistinguishable from interaction with $\RModel$. Finally, we will obtain the theorem by invoking \Cref{lemma:prf-switch}.

    Depending on whether $t$ is the first block in the response, we reason that $\calR_W = \calR_M$ differently:
    \begin{itemize}
        \item If~$r^{(t)}$ is the first block in the response, then $\calR_W = \calR_M$ because $\Wat^\calO$ and $\RModel$ behave identically until after the first block is sampled.
        \item If~$r^{(t)}$ is not the first block in the response (i.e., $r^{(t-1)}$ is a part of the same response as $r^{(t)}$), then $\calR_W = \calR_M$ because $r^{(t-1)} \not\in \{r^{(1)}, \dots, r^{(t-2)}\}$ and therefore $\Wat^\calO$ uses fresh randomness to generate $r^{(t)}$. (Note that $r^{(t-1)} \ne \bot$, since the subsequent block $r^{(t)}$ is a part of the same response.)
    \end{itemize}
    In either case, once we know that $\calR_W = \calR_M$, Inequality~\ref{ineq:substring-good} follows from the exact same argument as in the proof of \Cref{thm:wm_undet}.
\end{proof}





\section{Necessity of Assumptions} \label{sec:nec_assump}

In this section we show that the assumptions we use for our construction are necessary.
Informally, the two main statements we prove in this section are:
\begin{itemize}
    \item Undetectability is possible only against a computationally bounded adversary. That is, using a polynomial number of queries and exponential running time we can detect any nontrivial watermarking scheme. This is proven in \Cref{lem:comp_vs_stat}.

    \item Undetectability is impossible if outputs of low empirical entropy are watermarked: For any~$t$, if a non-negligible fraction of outputs with empirical entropy~$\leq t$ are watermarked then we can detect the watermark using~$\exp(t)$ queries and time. This is proven in \Cref{thm:nec-entropy}.
    
\end{itemize}
To warm up, we first observe that there exist models that generate text with arbitrarily high entropy, and nevertheless in any undetectable watermark only a negligible fraction of outputs can be watermarked. 
Hence, the natural attempt to only consider the model's entropy is insufficient. 

\begin{lemma}
    For every~$\lambda,b,\varepsilon>0$ there exists a prompt-independent model~$\Model$ such that~$H(\RModel(\emptyset))\geq b$, the maximum length of an output of~$\RModel$ is~$\frac{1}{\varepsilon} b$, and the following holds.
    If~$\calW$ is any undetectable and sound watermarking scheme for~$\Model$,
    then
    \[
    \Pr_{\substack{\sk \gets \Setup(1^\lambda) \\ x \gets \Wat_\sk(\emptyset)}}
    \left[\Detect_\sk \left(x\right)=\true\right]\leq \varepsilon + \negl
    .\]
\end{lemma}
\begin{proof}
    We define the output distribution of~$\RModel$ as follows. With probability~$1-\varepsilon$,~$\RModel$ outputs $\done$. Otherwise, it outputs a uniformly random binary string of length~$\frac{1}{\varepsilon}b$.
    The entropy of~$\RModel$ is larger than~$\varepsilon\cdot \frac{1}{\varepsilon}b > b$.
    
    Due to soundness, with high probability~$(\done)$ is not detected as watermarked, that is
    \[
    \Pr_{\sk \gets \Setup(1^\lambda)}
    \left[\Detect_\sk \left(\left(\done\right)\right)=\true\right]\leq \negl.
    \]
    On the other hand, due to undetectability,~$\Wat_\sk(\emptyset)$ must output~$(\done)$ with probability~$(1-\varepsilon)\pm \negl$, so the statement of the lemma holds.
\end{proof}

Next, we show that computational assumptions are necessary for undetectability of watermarks.
That is, while we can construct watermarks where the output distributions of~$\Wat_\sk$ and of~$\RModel$ are indistinguishable to any \emph{efficient} distinguisher, those distributions must not be identical and thus a distinguisher with unbounded running time is able to distinguish between them.
We prove a strong version of this statement: we show that for \emph{every} model and \emph{every} watermarking scheme it is possible to statistically distinguish~$\RModel$ from~$\Wat_\sk$, using a polynomial number of queries from any~$\prompt$ that produces watermarked outputs with non-negligible probability.

\begin{lemma}\label{lem:comp_vs_stat}
    Let~$\Model$ be a model and~$\calW$ a watermarking scheme for it that is sound.
    Let~$K$ be an upper bound on the size (in bits) of any possible secret key~$\sk$ for~$\calW$.
    Let~$\varepsilon > \frac{1}{\poly(\lambda)}$ and
    let~$\prompt$ be a prompt for which    
    \[
    \Pr_{\substack{\sk \gets \Setup(1^\lambda) \\ x \gets \Wat_\sk(\prompt)}}
    [\Detect_{\sk}(x) = \true] \geq \varepsilon
    .\]
    Then, for randomly chosen $\sk \gets \Setup(1^\secpar)$, it is possible to distinguish between~$\RModel$ and~$\Wat_\sk$ with probability at least~$\frac{1}{2}\varepsilon$, using~$\poly \left(\frac{K}{\varepsilon}\right)$ queries and~$\exp(K)$ running time.
\end{lemma}
\begin{proof}
    As~$\calW$ is sound, we in particular have that a random output is unlikely to be detected as watermarked,
    \[
    \Pr_{\substack{\sk \gets \Setup(1^\lambda) \\ x \gets \RModel(\prompt)}}
    [\Detect_{\sk}(x) = \true] \leq \negl
    .\]
    Let's call this property \emph{sound on average}.
    As soundness on average and the completeness assumption are distributional, we may assume that~$\Detect_\sk$ is deterministic by Yao's minimax principle, while maintaining both properties.\footnote{This means that we can always replace~$\Detect_\sk$ with a deterministic algorithm that still has those two properties. This is done by fixing the randomness used by~$\Detect_\sk$.}
    Therefore, for every possible secret key~$\sk$ there exists a subset~$W_\sk$ of possible outputs such that~$\Detect_\sk(x)=\true$ if and only if~$x\in W_\sk$.

    As~$\calW$ is sound on average, we have that
    \[
    \E_{\sk \gets \Setup(1^\lambda)}\left[
    \Pr_{x \gets \RModel(\prompt)}\left[x\in W_\sk\right] 
    \right]
    = 
    \Pr_{\substack{\sk \gets \Setup(1^\lambda) \\ x \gets \RModel(\prompt)}}
    [\Detect_{\sk}(x) = \true] \leq \negl.
    \]
    By the completeness assumption, we have
    \[
    \E_{\sk \gets \Setup(1^\lambda)} \left[
    \Pr_{x \gets \Wat_\sk(\prompt)}\left[x\in W_\sk\right]
    \right]
    =
        \Pr_{\substack{\sk \gets \Setup(1^\lambda) \\ x \gets \Wat_\sk(\prompt)}}[\Detect_\sk(x) = \true] \ge \varepsilon
    .
    \]
    In particular, due to both assumptions and Markov's inequality, with probability at least~$1-\frac{1}{2}\varepsilon$ the following hold for the drawn secret key~$\sk$:
    \begin{enumerate}
        \item $\Pr_{x \gets \RModel(\prompt)}\left[x\in W_\sk\right] < \frac{1}{2}\varepsilon^2$.
        \item $\Pr_{x \gets \Wat_\sk(\prompt)}\left[x\in W_\sk\right] > \varepsilon^2$.
    \end{enumerate}
    Let~$\calO$ be any distribution of strings, and~$W$ any set of strings. Let~$S$ be set of~$\frac{10}{\varepsilon^2}K$ independent strings drawn from $\calO$. 
    A Chernoff bound yields that
    \[
    \Pr_S\left[
    \Big|\; \frac{|S\cap W|}{|S|} - \Pr_{x \gets \calO}\left[x\in W\right]\; \Big| > \frac{1}{4}\varepsilon^2
    \right] \leq 3^{-K}
    .
    \]
    Given access to two distributions~$\calO_1,\calO_2$, we can draw~$\frac{10}{\varepsilon^2}K$ samples from each, denote them by~$S_1,S_2$ respectively, and then compute for every possible key~$k$ the quantities~$\frac{|S_1\cap W_k|}{|S_1|}$ and~$\frac{|S_2\cap W_k|}{|S_2|}$.
    Due to the Chernoff bound above and to a simple union bound, with probability at least~$1-2\cdot 1.5^{-K}$, those quantities approximate up to~$\pm \frac{1}{4}\varepsilon^2$ the probabilities~$\Pr_{x \gets \calO_j}\left[x\in W_k\right]$ for \emph{every} choice of possible key~$k$ and~$j\in\{1,2\}$.
    In that case, if~$\calO_1=\calO_2=\RModel$ there will not exist any~$k$ for which~$\frac{|S_1\cap W_k|}{|S_1|} < \frac{3}{4}\varepsilon^2$ and~$\frac{|S_2\cap W_k|}{|S_1|}>\frac{3}{4}\varepsilon^2$, but if~$\calO_1=\RModel$ and~$\calO_2=\Wat_\sk$, there would.
    Thus, we can distinguish between those two cases by making~$\frac{20}{\varepsilon^2}K$ queries and enumerating over all~$2^K$ possible keys.
\end{proof}

An important remark about Lemma~\ref{lem:comp_vs_stat} is that it requires the size of the secret key to be bounded (with respect to the number of queries the distinguisher makes).
If we wanted the distributions to be indistinguishable only to a much weaker distinguisher that is only allowed to sample each distribution once, or a bounded number of times, then we could achieve statistically identical distributions.
This can be achieved by the construction presented in Section~\ref{sec:wm}, if the PRF is replaced with a long secret key containing many random values that will each be used exactly once.

Finally, we show that any sound watermarking scheme that successfully watermarks outputs with empirical entropy smaller than~$t$, is detectable with~$\exp(t)$ queries and time.
This means that any watermarking scheme can only work for outputs with empirical entropy that is high enough with respect to the time the distinguisher is allowed to spend.

\begin{theorem} \label{thm:nec-entropy}
    Let~$\Model$ be a model and~$\calW$ a watermarking scheme for it that is sound with respect to a security parameter~$\lambda$. Let~$t$ be a parameter, and~$\prompt$ a prompt for which
    \[
        \Pr_{\substack{\sk \gets \Setup \\ x \gets \Wat_\sk(\prompt)}}[\Detect_\sk(x) = \true \text{ and }  \empH\left(\Model,\prompt,x\right) \leq t] > \frac{1}{\poly(\lambda)}.
    \]
    Then, it is possible to distinguish between the distributions~$\Wat_\sk$ and~$\RModel$ with~$O\left(\exp\left(t\right) \cdot \poly(\lambda)\right)$ queries and time, with a non-negligble probability.
\end{theorem}
\begin{proof}
    By making~$O\left(\exp\left(t\right) \cdot \poly(\lambda)\right)$ queries to a distribution~$\calO$ we may approximate with accuracy~$\pm 10^{-\lambda}$ the probability in the distribution~$\calO$ of every output~$x$ that has empirical entropy~$\leq t$. Note that there are at most~$2^t$ such outputs.
    By Lemma~\ref{lem:comp_vs_stat} and the assumption, there is a statistical difference between the distributions~$\RModel(\prompt)$ and~$\Wat_\sk(\prompt)$, even when we condition on the output being of empirical entropy~$\leq t$. 
    Hence, approximating both distributions on all such outputs is enough to distinguish between them. 
    Note that unlike in the proof of Lemma~\ref{lem:comp_vs_stat}, we are now not enumerating over all possible keys (that may be longer than~$t$ or~$\lambda$), we simply approximate both distributions and compute the statistical difference between them.
\end{proof}

\section{Removing Watermarks}\label{sec:remove}
A natural question is how robust an undetectable watermarking scheme can be to active attempts to remove it.
While we would ideally like to have an undetectable watermarking scheme that is robust to any efficient adversary attempting to remove a watermark, there are both practical and theoretical barriers to achieving this property.
In \Cref{subsec:attacks} we first describe several attacks that work well at removing watermarks in practice.
Then in \Cref{subsec:removing} we present an (expensive) attack that provably removes a watermark from any undetectable scheme.
We conclude that no undetectable watermarking scheme can be \emph{completely} unremovable. Still, it might require significantly more resources for a user to generate unwatermarked text from the model.

\subsection{Empirical Attacks on Watermarking Schemes} \label{subsec:attacks}

We highlight some relevant practical attacks, described for an attacker wishing to generate an unwatermarked response to a prompt $\prompt$. \cite{kgw} gives a nice overview of practical attacks removing watermarks, including a few of those mentioned here. 

\paragraph{Emoji attack.} 
In the ``emoji attack,'' the attacker asks the model to output a response to $\prompt$ with an emoji inserted between every pair of words.\footnote{\url{https://twitter.com/goodside/status/1610682909647671306}}
The attacker then removes the emojis to obtain the desired response.
This attack removes any watermark that relies on the detector seeing consecutive sequences of tokens, including ours as well as those of \cite{kgw} and \cite{scottblog}. In general this attack may not preserve the output distribution, but any provable robustness guarantee for contiguous-text watermarks would have to rest on the dubious assumption that it \emph{doesn't}.

\paragraph{Translation attack.}
The attacker can ask the model to write in a different language, and then translate the response to their language of choice. Depending on the fluency of the model in the other language, and the quality of the translation tool, this attack may significantly degrade the quality of text.

\paragraph{Paraphrasing and substitution attacks.}
The attacker obtains a response from the model. In the substitution attack, the attacker replaces some words with their synonyms. In the paraphrasing attack, the attacker paraphrases the text either manually or using a model as in \cite{krishna2023paraphrasing} or the span replacement attack of \cite{kgw}. Depending on how many words are changed, these attacks might remove the watermark from our scheme and those of \cite{kgw, scottblog}. Of course, changing more of the text also increases the risk of degrading its quality.


\paragraph{Post-hoc attacks.} For post-hoc detection schemes, there are two simple attacks that are empirically quite effective at evading detection. First, current LLMs are so powerful that they are often capable of simply evading detection themselves if you ask them to: See, for instance, this Reddit post.\footnote{\url{https://www.reddit.com/r/ChatGPT/comments/11pqmqm/you_can_ask_chat_gpt_to_write_a_text_than_alter/}.} Second, in some models, e.g. \href{https://platform.openai.com/playground}{OpenAI's ``Playground''}, one can change model parameters. By increasing the temperature, frequency penalty, and presence penalty, one can often produce text that evades post-hoc detection.


\subsection{Removing any Undetectable Watermark} \label{subsec:removing}
In this section, we describe an attack that removes a watermark from any undetectable watermarking scheme, assuming that the model is ``prefix-specifiable.'' The attack is simple: Just sample tokens from the watermarked model one at a time.

We say that a model is \emph{prefix-specifiable} if the user can specify a prefix of the model's response.
More formally, we require that for any $\prompt$ and text $x_1, \dots, x_k$, the user can efficiently compute a new prompt $\prompt'$ such that $\RModel(\prompt')$ is distributed identically to $\RModel(\prompt)$ conditioned on the response's prefix matching $(x_1, \ldots, x_k)$.
This property is also assumed in the definition of a language model in \cite{kgw}.

\paragraph{On prefix-specifiable models.}
Any language model according to \Cref{def:model,def:modelresp} can be given a prefix-specifiable interface, but real-world user interfaces may or may not allow it. For instance, ChatGPT does not allow the user to specify prefixes of the response, but the OpenAI Playground allows the user to submit text under the ``Assistant'' role which the model will use as a prefix for its next response. We do not know for certain if the resulting distribution is actually identical to the model's response conditioned on the given prefix.

We note that a user can always attempt to ``trick'' a model into being prefix-specifiable: Simply include in the prompt a request to start the response with the given prefix.
We also note that when we are defining a model, we can always design it to be ``prefix-specifiable'' by forcing it to follow such requests. In particular, a watermarking scheme that works for \emph{every} model would need to work also for prefix-specifiable models.


\begin{theorem}
    Let $\calW = (\Setup, \Wat_\sk, \Detect_\sk)$ be any undetectable watermarking scheme.
    Assume that the underlying model $\RModel$ is prefix-specifiable.
    Then there exists an efficient algorithm $\adv$ making queries to $\Wat_\sk$ such that, for any $\prompt$ and a random $\sk \from \Setup(\secparam)$, the distributions $\adv^{\Wat_\sk}(\prompt)$ and $\RModel(\prompt)$ are $\negl$-close in \emph{statistical} distance.
    The number of queries made by $\adv$ to $\Wat_\sk$ is exactly the length of text output by $\adv$.
\end{theorem}

\begin{proof}
The attacker generates an unwatermarked response to $\prompt$ as follows.
It lets $y_1$ be the first token of $\Wat_\sk(\prompt)$. It then lets $y_2$ be the first token of $\Wat_\sk(\prompt, y_1)$, and so on, in general computing $y_j$ as the first token in $\Wat_\sk(\prompt, y_1, \ldots, y_{j-1})$ until $y_j = \done$.

Let $Y = (Y_1, \ldots, Y_{L_Y})$ be random variables denoting the attacker's output.
Let $X = (X_1, \ldots, X_{L_X})$ be random variables denoting the output of $\RModel(\prompt)$; note that $L_Y$ and $L_X$ are random variables here.

We argue that for each $i$, $\Delta((Y_i \ | \ Y_{j} = y_j \ \forall j < i), (X_i \ | \ X_j = y_j \forall j < i)) \leq \negl$.
Suppose for the sake of contradiction that the total variation distance between these distributions for some $i$ is at least $\frac{1}{\poly(\secpar)}$ for some polynomial $\poly$.
A distinguisher trying to determine whether it has oracle access to $\RModel$ or $\Wat_\sk$ can make $O(\poly(|\calT|) \cdot \secpar)$ queries to the oracle with input $\prompt, (x_1, \dots, x_{i-1})$, and approximate the distribution of the first token in the response up to additive error $\pm 10^{-\secpar}$ in total variation distance. Since $10^{-\secpar} << 1/\poly(\lambda)$, this contradicts the undetectability of $\Wat_\sk$.


This tells us that for any partial response $(x_1, \ldots, x_i)$ to $\prompt$, the distribution of the next token must have $\negl$ total variation distance from that of the unwatermarked model.
Therefore, the entire response produced by this attack using the watermarked model must be have negligible total variation distance from the unwatermarked model.
\end{proof}

By soundness, the watermark cannot be detected in the output of $\adv$ with non-negligible probability.
Therefore, this attack succeeds in removing the watermark.

Fortunately, for reasonable text sizes this attack is probably impractical. For instance, while generating an $8,000$-token response from GPT-4 currently costs $\frac{\$0.06}{1000} \times 8000 = \$0.48$, generating the same text using the above attack would cost $\$0.48 + \frac{\$0.03}{1000} \times 8000 \times 7999 / 2 = \$960.36$. (Current GPT-4 pricing is $\$0.03$ per 1000 prompt tokens and $\$0.06$ per 1000 sampled tokens.)
Still, it shows that we cannot hope to achieve an overly sweeping definition of completeness/unremovability; it cannot allow an adversary powerful enough to run this attack. 

\section{Open Problems}\label{sec:open}
\Cref{sec:remove} implies that undetectable watermarking schemes cannot be made ``unremovable'' in a general sense.
Nevertheless, it is intriguing to find the most general sense in which undetectable watermarks can be made robust to removal attempts.
For example, our construction of substring-complete watermarks guarantees detection as long as a consecutive substring of the output with high enough empirical entropy remains intact.
Can this property be generalized to non-consecutive subsets of the output? Are there larger classes of \emph{removal techniques} against which undetectable watermarks can be made robust?

Quantitatively, our main schemes in \Cref{sec:wm} might not achieve an optimal completeness parameter. 
The simplified scheme we present in~\Cref{sec:inefficient} achieves completeness with a parameter that only depends on~$\lambda$, but not soundness or worst-case polynomial runtime. Our full construction in \Cref{sec:nec_assump}, on the other hand, achieves all required properties yet has a possibly non-optimal $\Theta\left(\lambda \sqrt{L}\right)$ completeness parameter.

Can we close this gap, either by improving our scheme to require less entropy for detection, or by showing a stronger lower bound for schemes that are sound (or efficient)?

\section*{Acknowledgments}
This research was supported in part by NSF Grants CCF-2107187, CCF-1763970, and CCF-2212233, by JPMorgan Chase \& Co, by LexisNexis Risk Solutions, and by the Algorand Centres of Excellence programme managed by Algorand Foundation. Any opinions, findings, and conclusions or recommendations expressed in this material are solely those of the authors.


\bibliographystyle{alpha}
\bibliography{sources}

\end{document}